\documentclass[%
 reprint,
 amsmath,amssymb,
 aps,
prb,
]{revtex4-2}
\usepackage{graphicx}
 \usepackage{filecontents}
\usepackage[utf8]{inputenc}
\usepackage[T1]{fontenc}
\usepackage{amsmath}
\usepackage{amssymb} 
\usepackage{bm}
\usepackage{physics}
\newcommand{\be}{\begin{equation}}
\newcommand{\ee}{\end{equation}}
\newcommand{\ba}{\begin{eqnarray}}
\newcommand{\ea}{\end{eqnarray}}
\newcommand{\bmat}{\begin{pmatrix}}
\newcommand{\emat}{\end{pmatrix}}
\usepackage{comment}
\usepackage{mathtools} 
\usepackage{braket}
\usepackage{esvect}
\usepackage{bbold}
\usepackage{palatino}
\usepackage{braket}
\usepackage{amsmath}
\usepackage{amssymb} 
\usepackage{graphicx}

\begin{document}

\title{Thermal rectification with topological edge states}

\author{Abdulla Alseiari}
\affiliation{Department of Physics, McGill University, 3600 rue University, Montreal, QC-H3A 2T8, Canada.}
\author{Michael Hilke}
\email{michael.hilke@mcgill.ca}
\affiliation{Department of Physics, McGill University, 3600 rue University, Montreal, QC-H3A 2T8, Canada.}

\begin{abstract}
Thermal rectification devices can be important for various thermal management applications. For oscillator chains, thermal rectification was observed when masses are distributed non-uniformly. Leaning on the importance of topological materials, we consider here a simple vibrational topological system, the binary isotope superlattice (BISL). We show that the BISL can be mapped exactly onto the Su-Schrieffer-Heeger (SSH) model, which has different topological phases, including topological edge sates. For the case, where there is a single topological edge state, we show that the BISL exhibits thermal rectification in the presence of a small nonlinear term. Thermal transport is computed using temperature reservoirs connected to both extremities. These results have implications for other classes of topological phonon systems. 
\end{abstract}

\maketitle

Thermal rectification is characterized by asymmetric heat conduction with respect to temperature gradients. It has emerged as a compelling area of study with promising applications in energy harvesting, thermal management, and nanoscale devices \cite{malik2022review,zhao2022review}. Following the pioneering work by Casati and co-workers \cite{terraneo2002controlling,li2004thermal,casati2007heat} a common strategy for thermal rectification is the use of graded materials, where the mass increases (or decreases) along the chain \cite{wang2012thermal}. A necessary ingredient for thermal rectification is the presence of nonlinear interactions and broken inversion symmetry \cite{chiu2016detecting}, so that the heat flux transmitted in one direction is different from the opposite direction. This asymmetry could be mediated by electrons, as in conventional diodes, or through phonons, which is the focus here. We are particularly interested in the case, where the electronic transport is symmetric with respect to a positive or negative applied potential difference, but not the heat flow. This is of interest for a variety of thermal devices \cite{schmotz2011thermal}. For instance, a thermal diode can be used as building block for thermal transistors, gates and memories \cite{li2006negative, wang2007thermal, wang2008thermal}. Thermoelectric diode structures could further improve the figure of merit in energy conversion \cite{hagelstein2002enhanced}.

In this work we closely follow the seminal framework by Lepri and co-workers \cite{lepri1997heat,lepri2003thermal} to compute the thermal conduction in one-dimensional lattices. In particular, we consider a chain with different masses $m_n$, but coupled by identical springs of spring constant $k=1$. In the harmonic limit, the Hamiltonian is simply given by 
\be H=\frac{\dot{x}_1^2}{2m_1}+\frac{1}{2}\sum_{n=2}^N \left(\frac{\dot{x}_n^2}{m_n}+(x_{n-1}-x_n)^2\right),
\label{Hamiltonian}
\ee
where $x_n$ is the deviation of mass $m_n$ from its equilibrium position. This represents an isolated harmonic chain with $N$ masses. The corresponding equation of motion for mass $n$ is then simply:
\be 
\ddot{x}_n=-\frac{1}{m_n}\frac{\partial H}{\partial x_n}.
\label{eqmotion}
\ee
The solution can be expressed using the standard dynamical matrix of size $N\times N$:

\be H_m=
\begin{pmatrix}
-\frac{1}{m_1}  & \frac{1}{m_1} & 0 & 0 & \cdots & 0\\
\frac{1}{m_2} & -\frac{2}{m_2}  & \frac{1}{m_2} & 0 & \dots & 0\\
\vdots & &  \ddots & & & \vdots\\
\cdots & \frac{1}{m_n}& -\frac{2}{m_n} & \frac{1}{m_n} & \cdots & 0\\
\vdots &  & & & \ddots & \vdots\\
  0& \cdots & 0 & 0  & \frac{1}{m_N}  & -\frac{1}{m_N} \\
\end{pmatrix},
\ee
which leads to a matrix form solution
\be H_m\vec{x}=\vec{\ddot{x}},\ee
where $\vec{x}$ is given by the elements $x_n$. It is now interesting to look at the corresponding strain solutions, where $s_n=x_{n+1}-x_n$ is the strain between masses $m_{n+1}$ and $m_n$. The strain equation now reads
\be H_s\vec{s}=\vec{\ddot{s}},\ee
where the $(N-1)\times (N-1)$ matrix $H_s$ is given by
\be H_s=
\begin{pmatrix}
-\frac{1}{m_1}-\frac{1}{m_2}& \frac{1}{m_2}& 0 & \cdots \\
\frac{1}{m_2} & -\frac{1}{m_2}-\frac{1}{m_3}& \frac{1}{m_3} & \cdots \\
\vdots & &  \ddots & \vdots\\
%\cdots & \frac{1}{m_n}& -\frac{1}{m_n}-\frac{1}{m_{n+1}}& \frac{1}{m_{n+1}} & \cdots & 0\\
%\vdots &  & & & \ddots & \vdots\\
\cdots & 0  & \frac{1}{m_{N-1}} & -\frac{1}{m_{N-1}}-\frac{1}{m_N} \\
\end{pmatrix}
\ee
and $\vec{s}$ are given by the elements $s_n$, where $n=1$ to $(N-1)$. 
The $(N-1)$ eigenvalues of the matrix $H_s$ are identical to the eigenvalues of the dynamical matrix $H_m$ with the exception of the zero eigenvalue of $H_m$,  which corresponds to a constant overall displacement and is absent in $H_s$.

We now consider the binary isotope superlattice (BISL), with $m_{2n+1}=1/t_2$ and $m_{2n}=1/t_1$, which yields $H_s=H_{SSH}-t_1-t_2$, where
\be
H_{SSH}=
\begin{pmatrix}
0& t_1& 0 & 0 & \cdots \\
t_1 & 0 & t_2 & 0 & \cdots \\
0 & t_2 & 0 & t_1 & \cdots \\
\vdots & & & \ddots &
%\cdots & 0  & \frac{1}{m_{N-1}} & -\frac{1}{m_{N-1}}-\frac{1}{m_N} \\
\end{pmatrix}.
\ee

\begin{figure}[h]
	\centering
		\includegraphics[scale=.5]{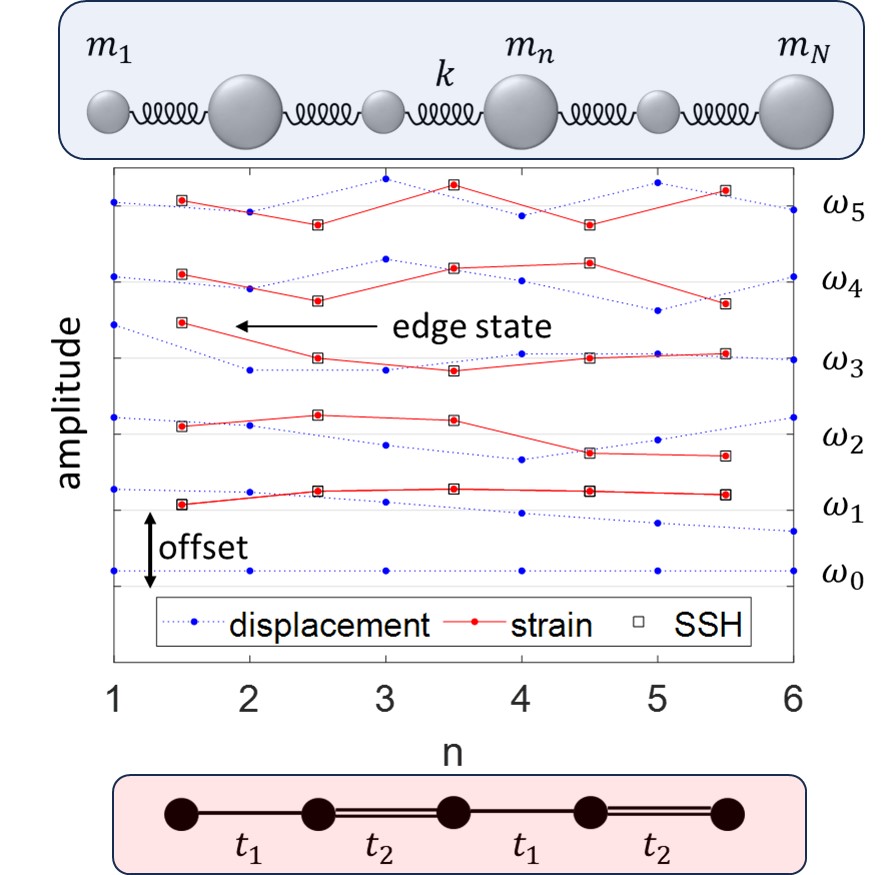}
	\caption{Top rectangle: illustration of the BISL with 6 alternating masses. Middle graph: the amplitude in blue of the 6 solutions of the BISL in terms of the displacement of each mass. In blue the 5 solutions corresponding to the strain between two masses and in black open squares the eigenvectors of the SSH model for 5 sites. We used the parameters $t_1=0.5$ and $t_2=2$ and offset each amplitude. Bottom rectangle: the illustration of the corresponding SSH model with alternating hopping elements. }
 \label{set-up}
\end{figure}

In $H_{SSH}$ we recognize the well known Su–Schrieffer–Heeger (SSH) model \cite{su1979solitons} for $(N-1)$ sites, where the hopping elements alternate between $t_1$ and $t_2$. Similarly to the oscillator system, the SSH model can be solved through the eigenvalue equation:
\be H_{SSH}\vec{\psi}^{(j)}=E_j \vec{\psi}^{(j)}, \ee
where $\vec{\psi}^{(j)}$ are the eigenvectors corresponding to the eigenvalues $E_j$. This equivalence between the SSH model and the BISL is illustrated in figure \ref{set-up}. The existence of gaps in biatomic lattices is well known \cite{hladky2007experimental} but not the connection to the SSH model.

The $N$-mass BISL, with masses $m_1,m_2,m_1,\cdots$ is equivalent to the $(N-1)$ sites SSH model with hopping elements $t_1,t_2,t_1,\cdots$, where $t_1=1/m_2$ and $t_2=1/m_1$. For $N-1$ odd, where $N-1$ is the number of sites in the SSH chain, and $t_1<t_2$ there will a single edge state localized on the left of the chain at zero energy, as illustrated in figure \ref{set-up}. In this case, the leftmost hopping element is $t_1$, while the rightmost hopping element is $t_2$. Correspondingly, for the BISL, the leftmost mass has a mass of $m_1=1/t_2$, i.e., $m_1<m_2$. This leads to the appearance of an edge displacement mode on the left as seen in figure \ref{set-up}, but none on the right, since the BISL is terminated by a mass $m_2>m_1$ for $N$ even. At frequency $\omega_3=\sqrt{1/m_1+1/m_2}$ the vibrations of the masses are localized on the left of the chain, where the strain $s$ follows the same spatial dependence as the solutions of the SSH model ($\vec{\psi}^{(j)}\sim\vec{s}^{(j)}$), where $E_j=t_1+t_2-\omega_j^2$ and ($j>0$). The oscillation frequencies of the BISL are given by the solution of  
\be H_m \vec{x}^{(j)}=-\omega_j^2\vec{x}^{(j)},\ee
where $\omega_j$ are the eigen-frequencies with $\omega_0=0$ ($\vec{x}^{(0)}$ does not depend on $n$) and $\vec{x}^{(j)}$
the eigenvectors. For $N=6$, $\omega_3=\sqrt{1/m_1+1/m_2}$ corresponds to $E=0$ of the SSH model. 

If we take the case $N$ odd, the BISL would terminate with masses $m_1$ on both extremities, which means that we would have 2 edge modes at $\omega\simeq \sqrt{1/m_1+1/m_2}$ and equivalently two edge states at $E\simeq 0$ in the SSH chain of size $(N-1)$, {\em i.e.}, of even length, with hoppings $t_1$ on both extremities. Here we assume $t_1<t_2$ and $m_1<m_2$. Note that there are 2 edge states for even sized SSH chains that are at energies close to zero but not exactly zero \cite{zaimi2021detecting}:
\be 
E = \pm t_2 (t_1/t_2)^{(N-1)/2}(1-t_1^2/t_2^2)+O(N-1).
\ee
This is referred to as the topological phase \cite{asboth2016short}. In the non-topological phase (or trivial phase), when $t_1>t_2$ there are no edge states but a gap instead. In this phase there are no solutions for $|E|<(t_1-t_2)$ \cite{su1979solitons}. Similarly, in the BISL, there will be a gap around $\omega\simeq \sqrt{1/m_1+1/m_2}$. 

The SSH model is a one-dimensional example of a topological insulator for electronic systems, where in the infinite size of the SSH chain, there is always a gap around $E=0$ and a continuum of states on each side of the gap. By making the chain finite, the number of edge states that appear inside the gap will depend on the topology. Equivalently, in two dimensions, as for example, in the integer quantum Hall effect, there is a gap given by the Landau levels, which closes for finite size, due to the appearance of chiral (topologoical) edge states that are responsible for the quantized Hall resistance \cite{klitzing1980new}. In fact most materials have a topological gap \cite{vergniory2022all}. %The topological phases can exist in all dimensions \cite{gao2015majorana,liu2019magnetic,zhang2010crossover} as well as in classical systems \cite{susstrunk2015observation,huber2016topological,vergniory2022all}.
While many topological materials are electronic in nature, they can also be observed in photonic and phononic systems \cite{RevModPhys.91.015006,liu2020topological}. Indeed, there are other phononic analogues of the SSH model, like with alternating springs \cite{chien2017thermal,chien2018topological,li2018schrieffer,liu2020topological} (but no discussion on rectification) or in photonics \cite{palmer2021asymptotically}. There are also topological phonon analogues in two dimensions \cite{wakao2020topological,xu2018topological}, including graphene \cite{li2020topological}. In two-dimensions, in analogy to the BISL, binary isotopic graphene superlattices have been studied experimentally for their phonon modes \cite{whiteway2020real,whiteway2020graphene} and thermal rectification has been evaluated for topological Mo\"bius graphene strips \cite{Jiang_2010}. 

To explore how phonon topology affects thermal transport and thermal rectification, we now look at the simple BISL case, where we demonstrated its topological nature above. Thermal transport in BISL can be obtained by weakly coupling a thermal Langevin reservoir at temperature $T_n$ to each mass $n$, with coupling $\lambda_n$. The equation of motion of mass $n$ due to the thermal reservoirs will be modified from equation \eqref{eqmotion} to \cite{lepri2003thermal}
\be 
\ddot{x}_n=-\frac{1}{m_n}\frac{\partial H}{\partial x_n}+\frac{\lambda_n}{m_n}\left(\epsilon_n\sqrt{2T_n/\lambda_n}-\dot{x}_n\right),
\label{eqmotion2}
\ee
where 
\be \langle\epsilon_n(t)\epsilon_m(t')\rangle=\delta_{n,m}\delta(t-t')\ee
are Gaussian stochastic variables. Using the fluctuation-dissipation theorem, the temperatures of the masses are given by 
\be T_n=m_n\langle \dot{x}_n^2\rangle
\label{T}
\ee 
in units of the Boltzmann constant ($k_B$) 
\cite{lepri2003thermal,dhar2008heat}. In what follows, we only couple our BISL to thermal reservoirs on the first and last masses, assuming $\lambda_{n}=0$ for $0<n<N$ and $\lambda_1=\lambda_N=\lambda$. The local heat current between masses $n$ and $(n+1)$ is given by \cite{lepri2003thermal,savin2014thermal}
\be 
J_{n+1/2}=-\frac{1}{2}\langle (\dot{x}_n+\dot{x}_{n+1})(x_{n+1}-x_n)\rangle.
\label{J}
\ee

\begin{figure}
    \centering
    \includegraphics[scale=.5]{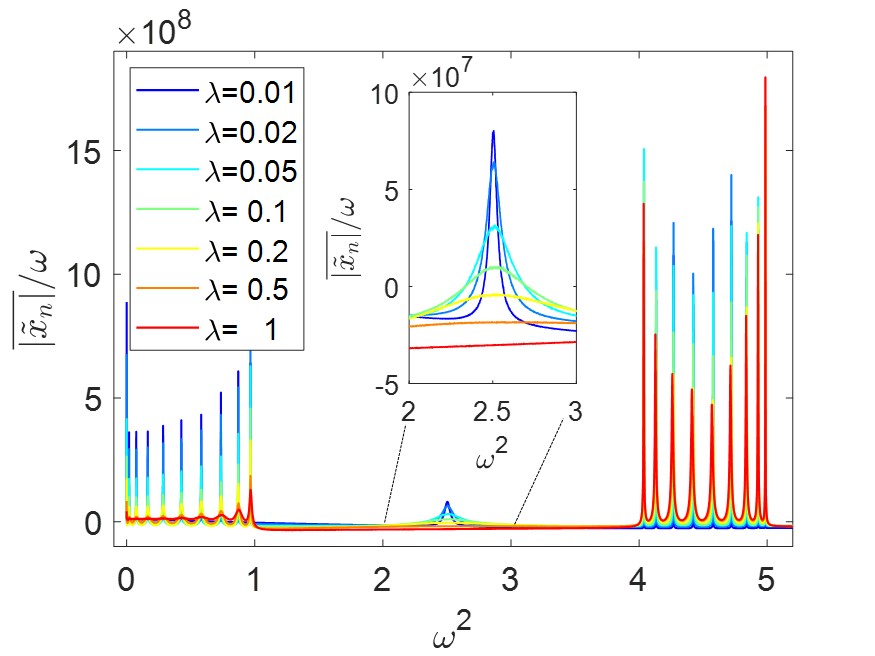}
    
    \includegraphics[scale=.5]{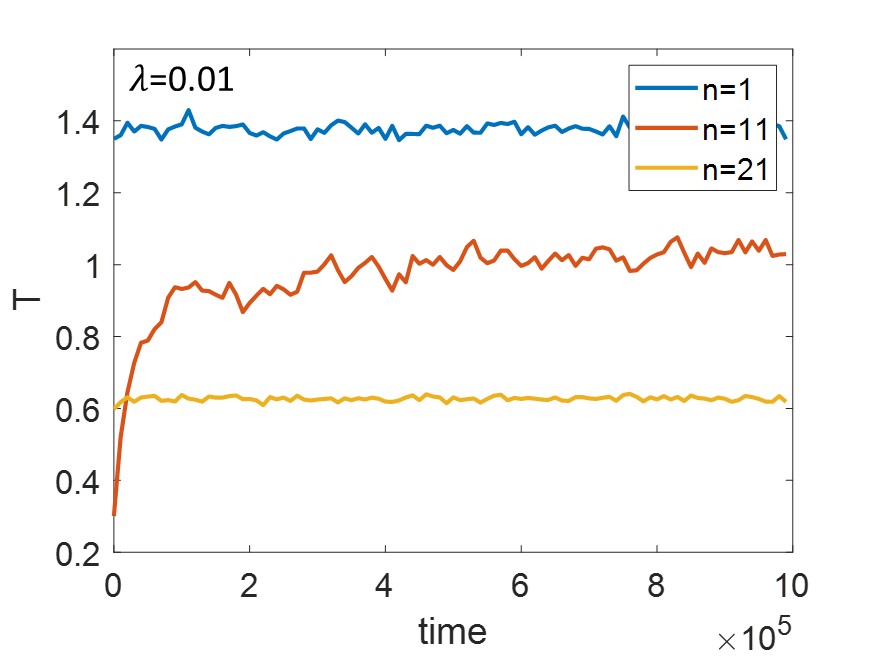}
    \caption{Top figure: Spectral density of the mass displacement. The average of the absolute value for all the masses is shown for different values of the coupling to the thermal reservoirs, $\lambda$. Bottom figure: The time dependence of the temperature for the first, middle and last masses. A chain of length $N=21$ is used here with $m_{2n+1}=0.5$ and $m_{2n}=2$ for $\lambda=0.01$. Each time data point is averaged over $2\times 10^5$ time steps and 100 different configurations. We used $T_1=1.5$ and $T_N=0.5$.}
    \label{lambda_dependence}
\end{figure}

We numerically solved equation \eqref{eqmotion2} using as time step $dt=0.05$ with all constants close or equal to one ($k_B=k=1$ and $0.5\leq m_l \leq 2$). Since the characteristic frequency is $1/\sqrt{m}$ we chose $dt/\sqrt{m}\ll 1$ for numerical accuracy. As we are interested in the properties of the free standing BISL, we further chose a weak coupling ($\lambda\ll 1$) to the heat reservoirs for the thermal rectification analysis below. However, it is important to make sure that we reach the steady state solution. A small coupling to the bath makes the thermalization between the reservoirs and the masses much slower. Hence, to make sure that we reach thermalization, a long enough time was used in the simulation as shown in fig. \ref{lambda_dependence}, where the temperature of 3 masses is shown as evaluated by using equ. \eqref{T}. The middle mass ($n=11$) has the slowest thermalization, but eventually saturates at time $T=10^6$. The number of iterations is given by $T/dt$.  

An important feature of SSH-equivalent systems is the existence of edge states in the gap. To visualize the spectrum, we can look at the spectral density $\left(\tilde{x}_n(\omega^2)\sim\int_0^T e^{-i\omega^2 t} x_n(t)\,dt\right)$ of our BISL system. The result is shown in fig. \ref{lambda_dependence} when averaged over $n$, with $T=10^6$. For larger values of the coupling $\lambda$, the broadening of the energy levels becomes quite substantial and we limit ourselves to  the case $\lambda=0.01$ for the rest of the paper. We note that in fig. \ref{lambda_dependence}, we see a peak structure of the spectral density in the gap close to $\omega^2=1/m_1+1/m_2=2.5$. This is due to the presence of edge states in the gap, which is a reflection of the case where the odd masses are lighter ($m_1=0.5$ and $m_2=2$). Reversing the values of $m_1$ and $m_2$ would yield no edge states in the gap.

\begin{figure}
    \centering
    \includegraphics[scale=.5]{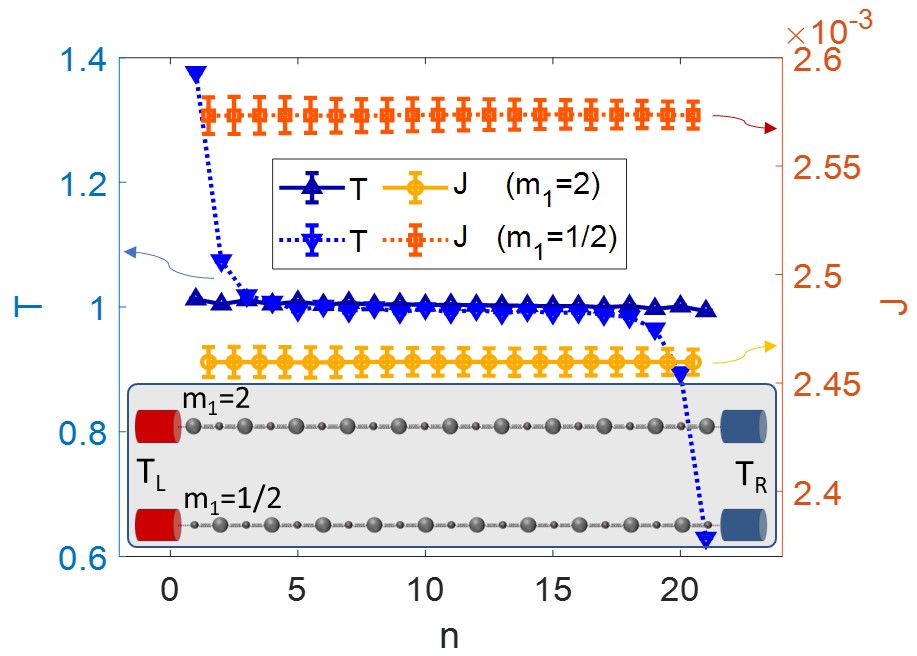}
\caption{Temperature dependence along the odd number chain ($N=21$) shown on the left scale. The corresponding heat current is shown on the right scale. The two possible configurations of the odd number chains are shown in the illustration. We used $\lambda_D=0.1$ for the nonlinear term and $lambda=0.01$. The rectangular inset illustrates the two configurations: $m_1>m_2$ (top) and $m_1<m_2$ (bottom). The error bars are standard errors obtained after averaging over 100 different time evolutions.}
    \label{temperature_dependence}
\end{figure}

In fig. \ref{temperature_dependence} we show the temperature dependence along the chain of the odd ($N=21$) case. We look at two cases: (1)  $m_1<m_2$ (topologically non-trivial with 2 edge states) and (2) $m_1>m_2$ (topologically trivial with zero edge states). In the first case we see a strong temperature dependence along the chain close to the edges, which is due to the existence of edge states. In the other case, where there are no edge states, the temperature is flat (independent of the mass position). We also computed the local heat current between two masses ($n$ and $n+1$) using equ. \eqref{J}. The heat current is slightly larger in the presence of edge states ($m_1<m_2$) as compared to the case without edge edge states ($m_1>m_2$). This is not very surprising since the presence of edge states in the gap will couple the BISL better to the thermal reservoirs then when there is only a gap. The edge states can lead to heat radiation \cite{ott2020radiative}. This behavior is similar to the electronic SSH model, where the electrical current is strongly suppressed for a Fermi energy at the band center in the case where no edge states are present \cite{zaimi2021detecting}. Inverting the temperatures of the heat reservoirs (not shown) does not change the absolute value of the heat current, only the sign, which is not surprising since the system is symmetric, hence no thermal rectification is observed in this case. Note, that we evaluate here the classical heat transport, which is different from the quantum heat transport, where a different phonon distribution needs to be used. Here we used  Langevin reservoirs which give a classical Poisson temperature distribution.

\begin{figure}
    \centering
    \includegraphics[scale=.5]{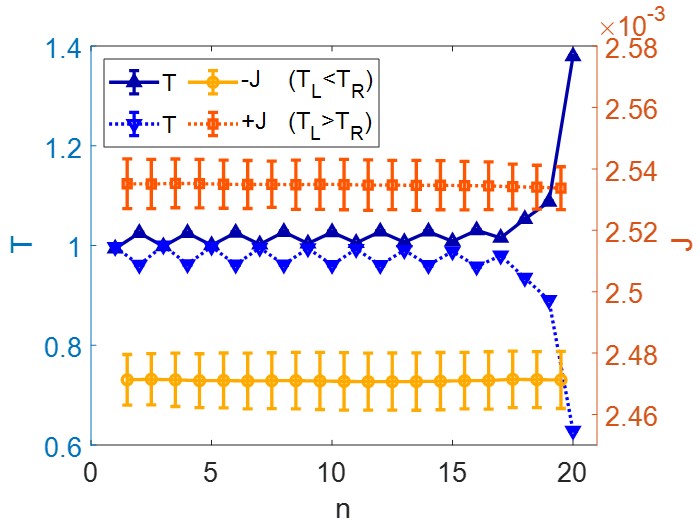}
    
    \includegraphics[scale=.5]{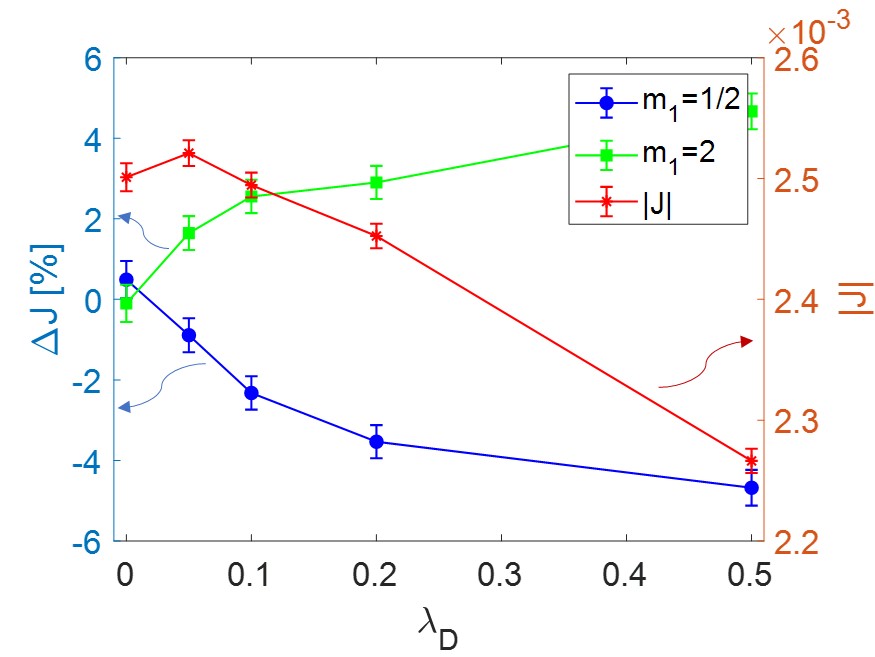}
    \caption{Top figure: Temperature dependence along the even number chain ($N=20$) for a non-linearity parameter $\lambda_D=0.1$ and with  $m_1=2$ and $m_2=0.5$. The absolute values of the heat currents are also shown for the two cases. Bottom figure: Dependence of the relative heat  rectification on $\lambda_D$ for the two cases ($m_1=0.5$ and $m_1=2$). The relative heat rectification is given by $\Delta J=2\frac{J(T_L>T_R)+J(T_L<T_R)}{J(T_L>T_R)-J(T_L<T_R)}$. The absolute value of the heat current is depicted as a function of $\lambda_D$ on the right scale.}
    \label{rectification}
\end{figure}

The even mass number case is very different and thermal rectification is obtained as shown below. This can be attributed to the inherent asymmetry in the system, since there will be one edge state either located on the left if $m_1<m_2$ or on the right if $m_1>m_2$. Moreover, a necessary ingredient for thermal rectification to occur was found to be a nonlinear potential in addition to the asymmetry \cite{terraneo2002controlling}. Here we consider the simplest non-linear term, which is an additional on-site potential term $V(x_n)=\lambda_Dx_n^4$ in the Hamiltonian (equ. \eqref{Hamiltonian}), where $\lambda_D$ is the strength of the non-linearity. The existence of nonlinear terms is very common in most materials \cite{cowley1968anharmonic,segal2003thermal,shah2013computer,zhu2021effects}.

We can now compute the temperature along the even mass BISL using this additional nonlinear term and the result is shown in fig. \ref{temperature_dependence}. Interestingly, in this case the temperature along the chain oscillates, which is different from the odd mass BISL (fig. \ref{temperature_dependence}). To look for rectification, we inverted the temperatures of the heat baths and observe a growing relative rectification with increasing non-linearity (see fig. \ref{rectification}). The absolute value of the heat current is larger for the case, where the lower temperature reservoir is on the side of the edge state and reduced in the opposite case. This means that the flow of heat out of the BISL is inhibited, when there is no edge state next to the cold reservoir. Consistently, inverting $m_1$ and $m_2$ switches the side where the absolute value of the heat current is larger, thus creating a directional heat switch by inverting $m_1/m_2$. We must note however, that the relative rectification is quite small (only a few percent). This is much smaller than other inhomogeneous mass systems, which for instance include long range interactions \cite{chen2015ingredients}. However, models that show much higher rectifications are very asymmetric, with typically a large change in on-site mass potential along the chain \cite{wang2012thermal,shah2012study,chen2015ingredients}. In BISLs, the asymmetry only occurs on one site at the edge, on an otherwise, uniform (or alternating) potential along the chain, which reduces the magnitude of rectification. Yet, the obtained rectification should be large enough to be observable in an experimental system, particularly in a two-dimensional analogue, where the topological edge would be spread over the entire width. Fundamentally, using topology as a means to induce thermal rectification adds a new dimension, not only for possible thermal rectification devices but also as a tool to probe topology. Indeed, by changing the topology, thermal rectification could be switched on or off, and vice-versa, topology could be measured via thermal rectification.  

Summarizing, we have shown that the peculiar edge state structure of an SSH-like binary mass system can lead to classical heat current rectification in the presence of a non-linear potential. We note that contrary to results in the literature, where a mass gradient along the chain is typically used, here we obtain heat rectification for an alternating binary mass distribution, which opens the door to finding heat rectification in another class of one-dimensional or higher dimensional phononic materials, which are topological in nature. 

We acknowledge useful discussions with Giulio Casati and Nic Delnour, and financial support from NSERC and INTRIQ.

\bibliography{refs}

\end{document}